\documentclass[tightenlines,eqsecnum,floats,aps,amsmath,amssymb%
,nofootinbib,prd,showpacs,twocolumn]{revtex4}

\usepackage{amsmath,amsthm,latexsym,amssymb,amsfonts}
\usepackage{enumerate}
\usepackage{graphicx}
\usepackage{calc}
\usepackage{colordvi}

\usepackage{LQC-symbols}

\newcommand{\cell}{\mathcal{V}}

\DeclareMathOperator*{\APS}{APS}
\DeclareMathOperator*{\sLQC}{sLQC}
\DeclareMathOperator*{\MMO}{MMO}

\newcommand{\blA}{\boldsymbol{A}}
\newcommand{\blB}{\boldsymbol{B}}
\newcommand{\blM}{\boldsymbol{M}}
\renewcommand{\id}{\openone}

\newcommand{\defi}{\mathcal{U}}

\begin{document}

\title{The LQC evolution operator of FRW universe with positive cosmological constant}

\author{Wojciech Kami\'nski${}^{1}$}
\email{wkaminsk@fuw.edu.pl}
\author{Tomasz Paw{\l}owski${}^{2,1}$}
\email{tomasz@iem.cfmac.csic.es}

\affiliation{
  ${}^{1}$Instytut Fizyki Teoretycznej, Uniwersytet Warszawski,
  ul. Ho\.{z}a 69, 00-681 Warszawa, Poland\\
  ${}^{2}$Instituto de Estructura de la Materia, CSIC,
  Serrano 121, 28006 Madrid, Spain
}

\begin{abstract}
  The self-adjointness of an evolution operator $\Theta_{\Lambda}$ corresponding
  to the model of flat FRW universe with massless scalar field and cosmological constant 
  quantized in the framework of Loop Quantum Cosmology is studied in the case $\Lambda>0$.
  It is shown, that for $\Lambda<\Lambda_c\approx 10.3\,\lPl^{-2}$ the operator admits many 
  self-adjoint extensions, each of the purely discrete spectrum. On the other hand
  for $\Lambda\geq\Lambda_c$ the operator is essentially self-adjoint, however the 
  physical Hilbert space of the model does not contain any physically interesting states.
\end{abstract}

\maketitle

\section{\label{sec:intro}Introduction}

Among various approaches to the unification of General Relativity and quantum physics
the Loop Quantum Gravity \cite{lqg,lqg-th} is one of the most promising. Its symmetry 
reduced version, Loop Quantum Cosmology \cite{lqc}, offers a qualitatively new picture 
of an early universe evolution \cite{aps-lett} and may provide a mechanism of solving 
long standing problems in modern cosmology \cite{awe-entr,ab-scinv}. However, 
although the number of works using the heuristic methods of mimicking the quantum evolution 
by an appropriately constructed classical mechanics \cite{eff} is rapidly growing, not so 
much effort has been dedicated so far to the investigation on a genuinely quantum 
level. The rigorous studies of these aspects are in fact restricted just to models 
either in vacuo \cite{mmp-b1,mmp-evo}, or admitting massless scalar field as the only 
matter content \cite{aps-imp,acs,mless,hiperb}. 
The number of works attempting to include the cosmological constant $\Lambda$ 
is even smaller \cite{boj-pos} and the rigorous analysis of the quantum 
universe dynamics within precise LQC model \cite{bp} was done only for negative 
$\Lambda$, a case not favored by the observations. 

This article is an attempt to partially fill this gap, by addressing the question 
whether, in presence of the \emph{positive} cosmological constant, the physical evolution 
defined by the methods currently applied in LQC \cite{aps-det} is unique. There, 
one treats the constrained system as a free one, evolving with respect to the scalar 
field regarded as an internal time. The evolution is generated by a so called evolution 
operator (further denoted as $\Theta$). On the technical level the definiteness and 
uniqueness of the evolution reduces to the existence and the uniqueness of the self-adjoint 
extensions of $\Theta$. In the previously investigated models this operator always 
admitted a unique extension \cite{kls-closed,kl-flat}, which ensured a unique evolution. 
The positive $\Lambda$ however acts like a negative unbounded potential, thus one can 
not immediately expect the same answer for the models with $\Lambda>0$. Here we 
analyze in detail the self-adjointness of $\Theta$, showing in particular, that for 
$\Lambda<\Lambda_c$, where $\Lambda_c$ is a certain critical value (of the Planck 
order) it in fact admits a family of extensions. This property is crucial for 
further studies of the universe dynamics \cite{ap-posL}

The paper is organized as follows. In section \ref{sec:frame} we briefly recall
the basic features of the model and introduce the elements relevant for 
our investigation. Next, in section \ref{sec:extensions} we determine the number 
of self-adjoint extensions  of $\Theta$ corresponding to, respectively, 
the \emph{subcritical} ($0<\Lambda<\Lambda_c$, Sec.~\ref{sec:subcr}) and \emph{supercritical}
($\Lambda\geq\Lambda_c$, Sec.~\ref{sec:supercr}) value of $\Lambda$, by probing 
the dimensionality of the deficiency spaces of $\Theta$ \cite{s-r} via the method 
presented in \cite{scmp} (Sec.~4). The properties of the physical Hilbert spaces
built of the spectral decomposition of $\Theta$ are briefly analyzed for both the 
subcritical and supercritical case in Secs.~\ref{sec:spectra} and \ref{sec:high}.
The article is concluded with Sec.~\ref{sec:concl}, where the results are summarized 
and their physical consequences as well as the direct extensions are briefly discussed.

\section{\label{sec:frame}The model}

Here we consider a model of a flat isotropic universe with positive cosmological 
constant $\Lambda>0$ and a free scalar field as a matter content (see 
Appendix A in \cite{aps-imp}). Its classical and kinematical description (in a loop 
quantization) is a direct analogy of the one used for the model with $\Lambda<0$ 
\cite{bp}. 

The considered spacetime admits a foliation (parametrized by a time $t$) by 
isotropic $3$-surfaces $\Sigma$ and the metric
\begin{equation}
  g = -N^2\rd t^2 + a^2(t) \fidq \ ,
\end{equation}
where $\fidq$ is a unit (fiducial) Cartesian metric on the surface $\Sigma$, $N$ 
is a lapse function and $a(t)$ is a scale factor. To describe the spacetime we 
use the canonical formalism, first selecting the fiducial triad $\fide^a_i$ orthonormal 
with respect to $\fidq$ (and the cotriad $\fidw^i_a$ dual to it), next introducing the 
canonical Ashtekar variables: connections $A^i_a$ and triads $E^a_i$, which upon 
partial fixing of the gauge freedom can be represented just by a pair of canonically 
conjugated variables: the connection $c$ and triad $p$ coefficients,
\begin{subequations}\label{eq:AE-def}\begin{align}
  A^i_a\ &=\ c V_o^{-\frac{1}{3}}\,\fidw^i_a \ ,  &
  E^a_i\ &=\ p V_o^{-\frac{2}{3}}\sqrt{\fidq}\,\fide^a_i \ ,
  \tag{\ref{eq:AE-def}}
\end{align}\end{subequations}
where $V_o$ is the fiducial (with respect to $\fidq$) volume of a certain comoving 
cubical region $\cell$ introduced to regulate the divergences of the action and 
the Hamiltonian. The coefficients $c$ and $p$ are global degrees of freedom.

Our system is a constrained one, with the diffeomorphism and Gauss constraints 
automatically satisfied by a gauge choice. The only nontrivial constraint is 
a Hamiltonian one
\begin{subequations}\label{eq:C-class}\begin{align}
  C &= N(C_{\gr} + C_{\phi}) , \\
  C_{\gr} &= -\frac{1}{\gamma^2} \int_{\fV} \rd^3 x \left( \varepsilon_{ijk}
             e^{-1} E^{ai} E^{bj} F^k_{ab} - \gamma^2\Lambda \right) , \\
  C_{\phi} &= 8\pi G p^{-\frac{3}{2}} p^2_{\phi} ,
\end{align}\end{subequations}
where $e:=\sqrt{|\det E|}$, $F^k_{ab}$ is the curvature of $A^i_a$:
$F^k_{ab}:=2\partial_a A^k_b + \varepsilon^k_{ij} A^i_a A^j_b$, $\gamma$ is the 
Barbero-Immirzi parameter, and $p_{\phi}$ is the canonical momentum of the scalar 
field $\phi$.

The quantization process is a direct application of the methods of LQG, following 
in particular the Dirac program consisting of the following steps: $(i)$ The system 
is first quantized on the kinematical level, with the constraint ignored. $(ii)$ Next 
the constraint is promoted to a quantum operator $\hat{C}$ defined in some domain 
of the kinematical Hilbert space $\Hilk$ identified in the previous step, and $(iii)$ 
the physical Hilbert space $\Hilp$ is built out of the states annihilated by it. Finally
$(iv)$ the evolution picture is provided by selecting an internal time (in our case 
this role is played by $\phi$) and defining the family of observables parametrized
by it. In a slightly weaker sense, the evolution is defined by the unitary mapping 
between the spaces of ``initial data''. In the cases considered here it corresponds 
to a map
\begin{equation}
  \re\ni\phi \mapsto \Psi(\cdot,\phi)\in\Hilk , \quad \Psi\in\Hilp.
\end{equation}
Our goal here is the verification of the existence and uniqueness of such mapping.

The particular realization and the results of each of theses steps is the following:

$(i)$ To assess the geometry degrees of freedom we construct the analog of an LQG 
holonomy-flux algebra consisting of the holonomies along the straight lines and fluxes 
along the unit square surfaces, then proceed with the quantization method used for LQG. 
The resulting kinematical gravitational Hilbert space is 
\begin{equation}
  \Hil^{\gr} = L^2(\bar{\re},\rd\mu_{\Bohr}) ,
\end{equation}
where $\bar{\re}$ is a Bohr compactification of the real line. 
The basic operators are holonomies $\hat{h}^{(\lambda)}$ and unit fluxes (or ``triads'') 
$\hat{p}$. A particularly convenient basis of $\Hil^{\gr}$ consists of the eigenstates 
of $\hat{p}$ labeled by $v\in\re$ as follows
\begin{equation}
  \hat{p}\ket{v} = (2\pi\gamma\lPl^2\sqrt{\Delta})^{2/3}\sgn(v)|v|^{2/3}\ket{v} ,
\end{equation}
where $\Delta$ is the LQC \emph{area gap} \cite{argap}.
In this basis the scalar product is given by
\begin{equation}\label{eq:ip-gr}
  \langle\psi|\psi'\rangle = \sum_{v\in\re} \bar{\psi}(v)\psi'(v) .
\end{equation}

The matter degrees of freedom are quantized via standard methods of quantum mechanics. 
In particular the basic operators are the field $\hat{\phi}$ and its momentum 
$\hat{p}_{\phi}$ and the matter Hilbert space is spanned by the eigenstates of $\hat{\phi}$.
The complete $\Hilk$ has thus the form
\begin{equation}
  \Hilk=\Hil^{\gr}\otimes\Hil^{\phi} , \quad \Hil^{\phi}=L^2(\re,\rd\phi) .
\end{equation}

$(ii)$ The constraint \eqref{eq:C-class} is first re-expressed in terms of holonomies 
and fluxes which are next promoted to operators. At present there are several prescriptions
existing in the literature, which differ in the technical details: choice of the lapse, 
factor ordering and symmetrization of an operator. In this paper we study three of them
introduced in \cite{aps-imp}, \cite{acs} and \cite{mmo} and denoted respectively by APS, 
sLQC and MMO prescriptions. In all of these cases the quantum constraint can be brought 
to the form
\begin{equation}\label{eq:constr-quant}
  \id\otimes\partial_{\phi}^2 + \Theta_{\Lambda} \otimes \id ,
  \qquad \Theta_{\Lambda} := \Theta_o - \Lambda V(v) ,
\end{equation}
where an action of the operator $\Theta_o$ equals
\begin{equation}\label{eq:ev2-gen}\begin{split}
  -[\Theta_o\psi](v) &= f_+(v)\psi(v-4) - f_o(v)\psi(v) \\ 
  &+ f_-(v)\psi(v+4) ,
\end{split}\end{equation}
with the form of $f_{o,\pm}$ depending on the particular prescription used and given 
respectively by
\begin{itemize}
  \item APS: 
    \begin{subequations}\label{eq:Theta-form-first}\begin{align}
      f_\pm(v) &= [B(v\pm 4)]^{-\frac{1}{2}}\tilde{f}(v\pm 2)[B(v)]^{-\frac{1}{2}} , \\
      f_o(v) &= [B(v)]^{-1} [f_+(v)+f_-(v)] , \\
      V(v) &= \pi G\gamma^2\Delta \frac{|v|}{B(v)},
    \end{align}\end{subequations}
    where \cite{rep-diff}
    \begin{subequations}\begin{align}
      \tilde{f}(v) &= (3\pi G/8) |v| \big| |v+1| - |v-1| \big| , \\
      B(v) &= (27/8) |v| \big| |v+1|^{1/3} - |v-1|^{1/3} \big|^3 . 
    \end{align}\end{subequations}
  \item sLQC: 
    \begin{subequations}\label{eq:fslqc}\begin{align}
      f_\pm(v) &= \frac{3\pi G}{4} \sqrt{v(v\pm 4)}  (v\pm 2) , \\  
      f_o(v) &= 3\pi G v^2 , \\
      V(v) &= \pi G\gamma^2\Delta\, v^2 .
    \end{align}\end{subequations}
  \item MMO:
    \begin{subequations}\label{eq:fmmo}\begin{align}
      f_{\pm}(v) &= C g(v\pm 4)s_{\pm}(v\pm 2)g^2(v\pm 2)s_{\pm}(v)g(v) , \notag \\
      f_o(v) &= C g^2(v) [ g^2(v-2)s^2_-(v) + g^2(v+2) s^2_+(v) ] , \notag \\
      V(v) &= \frac{8\pi G\gamma^2\Delta}{27}\frac{g^6(v)}{|v|} , \tag{\ref{eq:fmmo}}
    \end{align}\end{subequations}
    where
    \begin{subequations}\label{eq:Theta-form-last}\begin{align}
      g(v) &= \big| |1+1/v|^{1/3} - |1-1/v|^{1/3} \big|^{-1/2} , \\
      s_{\pm}(v) &= \sgn(v\pm 2) + \sgn(v) , \\
      C &= \pi G/12 .
    \end{align}\end{subequations}
\end{itemize}
In all the prescriptions the operators $\Theta_o$ and $\Theta_{\Lambda}$ are well 
defined in particular for $\varepsilon=0$ (see the detailed discussion in 
\cite{klp-aspects} for APS and \cite{mmo} for MMO).

$(iii)$ Given the constraint operator in the form \eqref{eq:constr-quant} one can 
find the physical Hilbert space for example by the group averaging techniques \cite{gave}.
For that however one needs to know the spectral decomposition of $\Theta_{\Lambda}$, thus its 
self-adjoint extension(s). 

Before going to this step let us note, that the structure of $\Theta_{\Lambda}$ 
and \eqref{eq:constr-quant} provides the natural division of the domain of $v$ onto the 
subsets (the \emph{lattices}) 
\begin{equation}
  \lat_{\varepsilon} = \{ \varepsilon+4n;\ n\in\integ\} ,\quad \varepsilon \in [0,4[
\end{equation}
preserved by the action of $\Theta_{\Lambda}$. This division is naturally transferred to the 
splitting of $\Hilp$ onto superselection sectors. In consequence it is enough 
to fix a particular value of $\varepsilon$ and work just with the restriction of the domain
of $\Theta_{\Lambda}$ to functions supported on $\lat_{\varepsilon}$ only. 

Further restriction comes from the fact that the considered system does not admit 
parity violating interactions. In consequence the triad orientation reflection 
$v\mapsto -v$, being a large gauge symmetry, provides another natural division onto superselection
sectors, namely the spaces of symmetric and antisymmetric states. For the rest of this work
we select the sector corresponding to symmetric states with $\varepsilon=0$. The studies
are however straightforward extendable to all other sectors, as we discuss 
at the end of each section. Our particular choice allows to further restrict the 
support of the functions to $\lat^+_{\varepsilon} := \lat^+_{\varepsilon} \cap \re^+$.

$(iv)$ A form of finding the physical Hilbert space and defining an evolution 
alternative to group averaging (and for the form of the constraint \eqref{eq:constr-quant}, 
equivalent to it \cite{GaSch-note}) is a reinterpretation of the system as a free one
evolving along the scalar field playing the role of an internal time. The similarity
of \eqref{eq:constr-quant} with the Klein-Gordon equation 
\begin{equation}\label{eq:constr-KG}
  [\partial^2_\phi\Psi](v,\phi) = -[\Theta_{\Lambda}\Psi](v,\phi)
\end{equation}
allows one to directly apply the standard quantum mechanical methods for solving it. 
Such structure in particular introduces yet another splitting onto superselection 
sectors corresponding to positive and negative energies out of which we select 
the positive sector. In consequence we can immediately write down the evolution between
the initial data states on the constancy surfaces of $\phi$, belonging to the projection
of $\Hil^{\gr}$ onto the space spanned by the positive part of the spectrum of $\Theta_{\Lambda}$
\begin{equation}\label{eq:Schroed}\begin{split}
  U_{\phi_o,\phi} : P_{\Theta_{\Lambda}\ge 0}\Hil^{\gr} &\to P_{\Theta_{\Lambda}\ge 0}\Hil^{\gr} , \\ 
  [U_{\phi_o,\phi}\Psi](v,\phi) &= e^{i(\phi-\phi_o)\sqrt{\Theta_{\Lambda}}_{+}} \Psi(v,\phi_o) ,
\end{split}\end{equation}
where the operator $P_{\Theta_{\Lambda}\ge 0}$ is the projection onto positive part of the spectrum 
and $\sqrt{\Theta_{\Lambda}}_{+}$ is the square root of $\Theta_{\Lambda}$ on the space 
$P_{\Theta_{\Lambda}\ge 0}\Hil^{\gr}$.
For the evolution to be well-defined and unitary however the operator $\Theta_{\Lambda}$ needs to 
be self-adjoint. Thus, the problem of the definiteness of the evolution reduces to 
the question about the self-adjointness of $\Theta_{\Lambda}$, which we will investigate in 
the next section.

\section{\label{sec:extensions}Extensions of the evolution operator}

To start with, we note that the operator $\Theta_{\Lambda}$ defined via \eqref{eq:constr-KG} 
and \eqref{eq:constr-quant} is symmetric on the domain $\Dom$ of the finite linear 
combinations of eigenstates $\ket{v}$ of $\hat{p}$, a set which is itself dense in 
$\Hil^{\gr}$. To check whether $\Theta_{\Lambda}$ is furthermore essentially self-adjoint we follow 
the method specified in \cite{s-r, scmp}: finding its deficiency indexes.

The first step is the identification of the deficiency subspaces $\defi^{\pm}$ defined
as the spaces of (kinematically) normalizable solutions $\psi^{\pm}$ to the equation
\begin{equation}\label{eq:defi-eq}
  [\Theta_{\Lambda}\psi^{\pm}](v) = \pm i\psi^{\pm}(v) .
\end{equation}
The dimensions of $\defi^{\pm}$ are exactly the deficiency indexes needed to verify 
the self-adjointness. By inspecting the form of $\Theta_{\Lambda}$ provided in 
\eqref{eq:constr-quant}-\eqref{eq:Theta-form-last} and taking into account the symmetry,
we note that any solution $\psi^{\pm}$ to \eqref{eq:defi-eq} is uniquely determined 
via its value $\psi^{\pm}(v=4)$. The spaces $\defi^{\pm}$ are thus at most $1$-dimensional
and nontrivial only when the solutions are normalizable.

To verify this property of $\psi^{\pm}$ we first analyze their asymptotes. 
To start with, we rewrite the equation \eqref{eq:defi-eq}, being the $2$nd order 
difference equation, in a $1$st order form, introducing:
\begin{equation}
  \vec{\psi}^{\pm}(v):= \left( \begin{array}{l} \psi^{\pm}(v) \\ \psi^{\pm}(v-4) \end{array} \right).
\end{equation}
In this notation the considered equation takes the form
\begin{equation}\label{eq:eig-1st}
  \vec{\psi}^{\pm}(v+4) = \blA(v)\,\vec{\psi}^{\pm}(v) ,
\end{equation}
where, applying the notation introduced in Eq.~\eqref{eq:constr-quant} and \eqref{eq:ev2-gen}, 
one can write the matrix $\blA$ as
\begin{equation}
  \blA(v) = \left( \begin{array}{cc} 
                     \frac{f_o(v)-\Lambda V(v)\mp i}{f_+(v)} & -\frac{f_-(v)}{f_+(v)} \\
                      1 & 0
                   \end{array} \right) .
\end{equation}
The next step is expressing $\vec{\psi}^{\pm}$ as a linear combination of the appropriately
selected asymptotic functions (further denoted as $\ub{\psi}^{\pm}_{\Lambda}$) and rewriting \eqref{eq:eig-1st} as 
the equation for the coefficients of that combination. At this point we note, that numerical inspection 
shows qualitatively different asymptotic behavior of $\psi^{\pm}$ depending on whether
the value of $\Lambda$ is below or above certain critical value $\Lambda_c$ related to 
the critical energy density $\rho_c$ \cite{aps-imp} or the area gap $\Delta$ as follows
\begin{equation}\label{eq:Lambda-c}
  \Lambda_c := 8\pi G\rho_c = 3/(\gamma^2\Delta) .
\end{equation}
Since the energy density operator has been shown to be bounded by $\rho_c$ 
\cite{acs,klp-aspects} and the cosmological constant carries a residual gravitational 
energy, it is natural from the physical point of view to restrict the consideration
just to $\Lambda<\Lambda_c$, although for completeness we will also dedicate 
some attention to the $\Lambda\geq\Lambda_c$ case.

\subsection{\label{sec:subcr}Subcritical $\Lambda$}

In this case, as asymptotic functions we select 
$\ub{\psi}^{\pm}_{\Lambda}$ defined as
\begin{equation}\label{eq:defi-as-def}
  \ub{\psi}^{\pm}_{\Lambda} := |v|^{-1} e^{\pm i\omega(\Lambda)|v|} ,
\end{equation}
where 
\begin{equation}
  \omega(\Lambda) = \frac{1}{2}\arccos(\sqrt{\Lambda/\Lambda_c}) .
\end{equation}
With that choice we define the vector of coefficients for each pair of consecutive 
points on $\lat^+_0$
\begin{equation}\label{eq:eig-coeff}
  \vec{\psi}^{\pm}(v) = \blB(v-4) \vec{\chi}^{\pm}(v) ,
\end{equation}
where the transformation matrix $\blB$ is defined as 
\begin{equation}
  \blB(v) := \left( \begin{array}{ll} 
                      \ub{\psi}^+_{\Lambda}(v+4) & \ub{\psi}^-_{\Lambda}(v+4) \\
                      \ub{\psi}^+_{\Lambda}(v) & \ub{\psi}^-_{\Lambda}(v)
                    \end{array} \right) .
\end{equation}
Having that, we can rewrite the equation \eqref{eq:eig-1st} as follows
\begin{equation}\label{eq:M-def}\begin{split}
  \vec{\chi}^{\pm}(v+4) &= \blB^{-1}(v) \blA(v) \blB(v-4) \vec{\chi}^{\pm}(v) \\
  &=: \blM(v) \vec{\chi}^{\pm}(v) .
\end{split}\end{equation}
The exact coefficients of the matrix $\blM(v)$ can be calculated explicitly. 
The property relevant for us is that for each of the prescriptions listed in 
Sec.~\ref{sec:frame} it features the following asymptotic behavior
\begin{equation}
  \blM(v) = \id + \boldsymbol{O}(v^{-2}) , 
\end{equation}
where $\boldsymbol{O}(v^{-n})$ denotes a matrix, whose coefficients 
asymptotically behave as $O(v^{-n})$. This implies immediately (see Sec.~$4$ 
of \cite{scmp}) the existence of the limit 
\begin{equation}\label{eq:defi-lim-def}
  \lim_{n\to \infty} \vec{\chi}^{\pm}(4n) =: \vec{\chi}^{\pm} ,
\end{equation}
such that
\begin{equation}\label{eq:coeff-asympt}
  \vec{\chi}^{\pm}(v) = \vec{\chi}^{\pm} + \vec{O}(v^{-1}) .
\end{equation}
In consequence
\begin{equation}\label{eq:defic-limit}
  \psi^{\pm}(v) 
  = (\ub{\psi}^+_{\Lambda}(v),\ub{\psi}^-_{\Lambda}(v)) \cdot \vec{\chi}^{\pm}
  + O(v^{-2}) .
\end{equation}
This, together with the fact, that $\psi^{\pm}$ is well defined and finite everywhere,
implies that their norm with respect to the inner product \eqref{eq:ip-gr} is finite. 

Combining the above observation with the structure of eigenspaces discussed earlier 
we conclude, that the deficiency spaces $\defi^{\pm}$ are both $1$-dimensional. Therefore
\cite{s-r,scmp} the operator $\Theta_{\Lambda}$ \emph{is not} essentially self-adjoint, although 
it \emph{admits a family} of self-adjoint extensions. Each extension corresponds to the
unitary transformation 
\begin{equation}\label{eq:utr-def}
  U:\defi^+ \to \defi^- .
\end{equation}
Since $\defi^{\pm}$ are $1$-dimensional the only possible transformations which map the normalized 
$\psi^+$ into the space $\defi^-$ are as follows
\begin{equation}\label{eq:U-form-1d}
  \psi^+ \mapsto U^{\alpha}\psi^+ = e^{i\alpha}\psi^- ,
\end{equation}
where $\psi^-$ is also assumed to be normalized. The family of possible extensions is thus
labeled by one parameter $\alpha\in [0,2\pi]$.

The above result can be extended in a straightforward way to other superselection sectors 
labeled by $\varepsilon$. The particular form of the extension and the detailed of 
its result depend on the prescription used. For the MMO one, since the triad 
orientations (positive and negative $v$) separate (see the detailed discussion in
\cite{mmp-b1,mmo}) one can always restrict the consideration to $v>0$. In consequence
the space of solutions to \eqref{eq:defi-eq} is again $1$-dimensional and the deficiency 
functions are uniquely determined by their value at $v=\varepsilon$. Thus, the 
analysis of the asymptotics described above can be applied in this case without 
any modifications providing exactly the same result as for $\varepsilon=0$.

For the remaining two prescriptions the situation is slightly more complicated. Namely,
for generic $\varepsilon$ the eigenspaces of $\Theta_{\Lambda}$ corresponding to 
any eigenvalue, including $\pm i$, are $2$-dimensional. Also to verify their normalizability
one needs to check the asymptotics independently for positive and negative $v$. 
Nonetheless it can still be done by direct application of the method used for 
$\varepsilon=0$ to each of the limits. The result is the same, although while analogs 
of \eqref{eq:defi-lim-def} are still well defined (and the rate of convergence is the same),
generically
\begin{equation}
  \lim_{n\to\infty} \vec{\chi}(\varepsilon+4n) \neq \lim_{n\to\infty} \vec{\chi}(\varepsilon-4n).
\end{equation}
As for $\varepsilon=0$ all the solutions to \eqref{eq:defi-eq} are normalizable. Now however
$\dim(\defi^+) = \dim(\defi^-) = 2$, so the self-adjoint extensions of $\Theta_{\Lambda}$ 
are now labeled by the elements of the $U(2)$ group. 

Restricting the studies to the symmetric functions does not change the result for 
$\varepsilon\neq 2$ as the parity reflection maps the lattice $\lat_{\varepsilon}$ onto
$\lat_{4-\varepsilon}$, disjoint from the original one. In the only exceptional case
$\varepsilon=2$, the symmetry imposes an additional constraint between the values of 
the eigenfunctions at $v=2$ and $v=-2$. In consequence the eigenspaces are again 
$1$-dimensional and the results are exactly the same as for $\varepsilon=0$.

\subsection{\label{sec:supercr}Supercritical $\Lambda$}

In the case $\Lambda\geq \Lambda_c$ it is convenient to introduce the following 
change of representation for a general superselection sector $\varepsilon$
\begin{equation}\label{eq:super-map}
  \psi(v) \mapsto \tilde{\psi}(v) = (-1)^{(v-\varepsilon)/4}\psi(v) .
\end{equation}
It is trivial to note that the kinematical inner product \eqref{eq:ip-gr} between 
transformed functions is given by a formula identical to \eqref{eq:ip-gr}. 
On the other hand the examination of the form of $\Theta_{\Lambda}$ provided by
\eqref{eq:constr-quant}-\eqref{eq:Theta-form-last} shows that it transforms into
\begin{equation}\label{eq:super-trans}
  \Theta_{\Lambda} \to \tilde{\Theta}_{\Lambda} = -\Theta_{\Lambda_c-\Lambda} + A(v)\id ,
\end{equation}
where $A(v)$ is always finite and decays as $O(v^{-2})$, thus $A(v)\id$ is a compact 
operator. This feature allows immediately to apply Kato's perturbation theory 
\cite{kato} and the self-adjointness of $\Theta_{\Lambda}$ for $\Lambda\leq 0$ 
\cite{kl-flat} to conclude, that for $\Lambda\geq\Lambda_c$ the operator 
$\Theta_{\Lambda}$ is also essentially self-adjoint.

\section{\label{sec:spectra}The spectral properties for $\Lambda<\Lambda_c$}

For subcritical values of $\Lambda$ we have shown in Sec.~\ref{sec:subcr} that 
the evolution operator $\Theta_{\Lambda}$ admits (in the principal case $\varepsilon=0$ 
considered here) $1$-parameter family of extensions $\Theta_{\alpha}$. Each of 
these extensions defines an evolution via \eqref{eq:Schroed} with $\Theta_{\Lambda}$ replaced 
by $\Theta_{\alpha}$. In order to identify the physical Hilbert space 
$\Hil^{\phy}_{\alpha}$ corresponding to each extension we need to know (the positive 
part of) the spectrum of $\Theta_{\alpha}$. Here we analyze some properties of it, 
as well as the eigenspaces corresponding to its elements.

Let us start with the eigenfunctions. By inspection one can easily notice that 
the analysis of the asymptotics of the deficiency functions performed in Sec.~\ref{sec:extensions}
extends directly to any eigenfunction corresponding to any complex eigenvalue, with
the same form of the asymptotic functions \eqref{eq:defi-as-def} and convergence 
rates \eqref{eq:coeff-asympt}. In consequence every eigenfunction of $\Theta_{\Lambda}$ is 
explicitly normalizable, being thus an element of $\Hil^{\gr}$. This in turn implies
that the spectrum of each $\Theta_{\alpha}$ is purely discrete. 

To identify the spectra $\Sp(\Theta_{\alpha})$ we first determine the domain of 
each extension, applying the theorem X.2 of \cite{s-r}. It follows from it, that 
the domain $\Dom_{\alpha}$ of $\Theta_{\alpha}$ in our case equals to
\begin{equation}\label{eq:dom-ext}
  \Dom_{\alpha} = \{ \psi + \psi^{+} + U^{\alpha}\psi^{+} ;\ 
  \psi\in\Dom,\, \psi^{\pm}\in\defi^{\pm} \} ,
\end{equation}
where $U^{\alpha}$ is given by \eqref{eq:U-form-1d}. On the other hand $\Dom_{\alpha}$
is spanned by those of the (normalized) eigenfunctions $e_{\omega}(v)$ whose eigenvalues
$\omega\in\Sp(\Theta_{\alpha})$. As the eigenfunctions are normalizable, the ones selected 
by that condition also belong to $\Dom_{\alpha}$. Since the original domain $\Dom$ of 
$\Theta_{\Lambda}$ is (a Cauchy completion with respect to the graph norm of) a space of 
finite linear combinations of $\ket{v}$, only the term $\psi^{+} + U^{\alpha}\psi^{+}$ 
contributes to the asymptotics of the elements $\Dom_{\alpha}$. In consequence $\Dom_{\alpha}$ 
is spanned by (all and only) the eigenfunctions $e_{\omega}$ which converge to a combination
$\psi^{+} + U^{\alpha}\psi^{+}$ for some $\psi^+\in\defi^+$.

% altermative form of the asymptotics
The above selection criterion, although precise, is not convenient for practical 
purposes. To bring it to a simpler form, we remind that all the eigenfunctions, 
including the deficiency functions and $e_{\omega}$, converge to linear combinations
of $\ub{\psi}^{\pm}_{\Lambda}$. Furthermore, as $\Theta_{\Lambda}$ is a real operator, the 
limit of $e_{\omega}$ is necessarily of the form
\begin{equation}\label{eq:beta-lim}\begin{split}
  e_{\omega}(v) &= \lambda(\omega)\left[ e^{i\beta(\omega)}\ub{\psi}^+_{\Lambda}(v)
     + e^{-i\beta(\omega)}\ub{\psi}^-_{\Lambda}(v) \right] \\ 
  &+ O(v^{-2}),
\end{split}\end{equation}
where $\lambda(\omega)\in\compl$ and the phase shifts $\beta(\omega)\in[0,2\pi]$. Obviously the
term $\psi^{+} + U^{\alpha}\psi^{-}$ has the same form of the limit, up to an additional 
rotation by a global phase. Furthermore, the transformation $\beta\to\beta\pm\pi$ 
corresponds just to change of sign. In consequence there is a one to one correspondence 
between the parameters $\alpha$ and $\beta\in[0,\pi[$ which thus uniquely label 
the extensions.

As one needs just to compare the asymptotic behavior of the eigenfunction against 
the functions of a very simple analytic form, the classification with respect to 
$\beta$ is much better suited for practical applications, like e.g. the explicit identification 
of the spectra of the extended operators, as well as for finding the bases of the physical 
Hilbert spaces. One has to remember however, that this classification is just a more 
convenient form of the previous one, not an alternative to it.

% comment on other sectors
The above results, derived for the superselection sector $\varepsilon=0$, generalize 
easily to other sectors, although the exact results depend (as in the studies of 
Sec.~\ref{sec:subcr}) on the particular prescription. Namely, for the MMO prescription, 
due to nondegeneracy of the eigenspaces of $\Theta_{\Lambda}$, the analysis presented in this 
section can be repeated exactly, giving exactly the same results. For the remaining 
two prescriptions one has to introduce slight modifications taking into account the 
twofold degeneracy of the eigenvalues. In particular the label of the extension, 
inherited from the label of the unitary transformation $U$ \eqref{eq:utr-def} via
\eqref{eq:dom-ext} is now an element of the $U(2)$ group. All the eigenfunctions of 
$\Theta_{\Lambda}$ are however again explicitly normalizable, and the ones spanning a particular
extension are selected by the condition that a given eigenfunction $e_{\omega}$ belongs to
$\Dom_{\alpha}$ iff there exists $\psi^+\in\defi^+$ such that the considered eigenfunction 
converges to a combination $\psi^++U^{\alpha}\psi^+$, where $U^{\alpha}$ is a transformation 
\eqref{eq:utr-def} corresponding to a particular value of the label $\alpha\in U(2)$.

\section{\label{sec:high}Physical Hilbert space for $\Lambda\geq\Lambda_c$}

For these cases we have proved in Sec.~\ref{sec:supercr} that the operator $\Theta_{\Lambda}$
is essentially self-adjoint. Also the form \eqref{eq:super-trans} of $\Theta_{\Lambda}$ after 
the representation change \eqref{eq:super-map} suggests, that qualitatively its 
spectrum should resemble $\Sp(\Theta_{\Lambda'})$, where $\Lambda'=\Lambda_c-\Lambda\leq 0$.
Thus we expect the whole spectrum to be quite rich. In particular the essential part
or it equals just $\Sp_{\rm es}(-\Theta_{\Lambda'})$. As $\Sp_{\rm es}(\Theta_{\Lambda'})$
equals either $\re^+$ (for $\Lambda'=0$) or is empty ($\Lambda'<0$) \cite{kl-flat} 
$\Sp_{\rm es}(\Theta_{\Lambda})$ is purely nonpositive. On the other hand, since 
only $P_{\Theta_{\Lambda}>0}\Hil^{\rm gr}$ enters \eqref{eq:Schroed}, only the positive 
part of $\Sp(\Theta_{\Lambda})$ 
is relevant from the physical point of view. From the above reasoning it follows 
immediately that it has to be purely discrete \cite{cont-note}. In this section we will study exactly 
this part. The analysis will be again restricted just to the superselection 
$\varepsilon=0$ and the symmetric functions.

The discreteness of the positive part of the spectrum implies, that the eigenfunctions 
corresponding to it have to be explicitly normalizable. One can show however, that 
no such function exists at least for selected prescriptions and superselection sector.
Indeed, from \eqref{eq:constr-quant},~\eqref{eq:ev2-gen} and the positivity of 
$f_{o,\pm}(v)$ and $\Lambda_c V(v)-f_o(v)$ for $v\geq 4$ follows, that the solution to
the equation $\Theta_{\Lambda}\psi_{\omega,\Lambda} = \omega^2\psi_{\omega,\Lambda}$ 
satisfies the relation
\begin{subequations}\label{eq:eigenf-neq}\begin{align}
  |&\psi_{\omega,\Lambda}(v+4)| \notag\\
  &\hphantom{\psi}\geq \frac{\Lambda V(v)-f_o(v)+\omega^2}{f_+(v)}|\psi_{\omega,\Lambda}(v)|
  - \frac{f_-(v)}{f_+(v)}|\psi_{\omega,\Lambda}(v-4)| \tag{\ref{eq:eigenf-neq}} \\
  &\hphantom{\psi}\geq \frac{\Lambda_c V(v)-f_o(v)}{f_+(v)}|\psi_{\omega,\Lambda}(v)|
  - \frac{f_-(v)}{f_+(v)}|\psi_{\omega,\Lambda}(v-4)| . \notag
\end{align}\end{subequations}
This and the fact that for chosen superselection sector all the eigenfunctions 
$\psi_{\omega}$ are determined by their initial values $\psi_{\omega}(v=4)$ implies
\begin{equation}
  |\psi_{\omega,\Lambda}(4)| = |\psi_{0,\Lambda_c}(4)|\  
  \Rightarrow\
  |\psi_{\omega,\Lambda}(8)| \geq |\psi_{0,\Lambda_c}(8)| ,
\end{equation}
thus, defining the ratios 
\begin{equation}
  \chi_{\omega,\Lambda}(v) := - \psi_{\omega,\Lambda}(v)/\psi_{\omega,\Lambda}(v-4) ,   
\end{equation}
we have 
\begin{equation}
  \chi_{\omega,\Lambda}(8) \geq \chi_{0,\Lambda_c}(8).  
\end{equation}
Furthermore, $\chi_{\omega,\Lambda}$ satisfy the equation (following directly 
from \eqref{eq:constr-quant},~\eqref{eq:ev2-gen})
\begin{equation}
  \chi_{\omega,\Lambda}(v+4) = \frac{\Lambda V(v)-f_o(v)+\omega^2}{f_+(v)}
  - \frac{f_-(v)}{f_+(v)} \frac{1}{\chi_{\omega,\Lambda}(v)} ,
\end{equation}
which together with (following from \eqref{eq:Theta-form-first}-\eqref{eq:Theta-form-last})
positivity of $f_{\pm}(v)$ and $\Lambda_c V(v)-f_o(v)$ for $v\geq 4$ implies
\begin{equation}\begin{split}
  \forall v\geq 8 :\ \  
  &\chi_{\omega,\Lambda}(v) \geq \chi_{0,\Lambda_c}(v) \\ &\Rightarrow\ \ 
  \chi_{\omega,\Lambda}(v+4) \geq \chi_{0,\Lambda_c}(v+4) .
\end{split}\end{equation}
In consequence, by induction we have
\begin{equation}\label{eq:eigenf-growth}\begin{split}
  &|\psi_{\omega,\Lambda}(4)| \geq |\psi_{0,\Lambda_c}(4)| \ \ \Rightarrow \\
  &\forall n\in\integ^+:\ |\psi_{\omega,\Lambda}(4n)| \geq |\psi_{0,\Lambda_c}(4n)| .
\end{split}\end{equation}
On the other hand, taking as $\ub{\psi}^\pm(v)$ the functions
\begin{equation}
 \ub{\psi}^+(v)=\frac{(-1)^{v/4}}{\sqrt{|v|}},\ \ 
 \ub{\psi}^-(v)=\frac{(-1)^{v/4}}{\sqrt{|v|}}\ln|v|,
\end{equation}
one can perform the analysis of the asymptotics analogous to 
the one in Sec.~\ref{sec:subcr}, showing that 
\begin{equation}\label{eq:0c-limit}
  \psi_{0,\Lambda_c}(v) = \frac{(-1)^{v/4}}{\sqrt{|v|}} (c_1  + c_2\ln|v|) 
  + O(v^{-3/2}\ln(v)) ,
\end{equation}
where, due to the existence of both $\prod_{n=n_0}^\infty M(4n)$ and 
$\prod_{n=\infty}^{n_0} M(4n)^{-1}$ (where $M$ is an analog of the matrix defined in
\eqref{eq:M-def}) 
for some large enough $n_0$, and the fact that the eigenfunction is uniquely determined 
by its value at $v=4$, at least one of the coefficients $c_1$, $c_2$ does not vanish.

From the relations \eqref{eq:eigenf-growth} and \eqref{eq:0c-limit} we see, that for
$\Lambda\geq\Lambda_c$ none of the eigenfunctions corresponding to the positive 
eigenvalues are normalizable. In consequence the positive part of the spectrum
of $\Theta_{\Lambda}$ is empty, thus the physical Hilbert spaces corresponding to 
those values of $\Lambda$ are trivial.

This result cannot be immediately extended to the remaining superselection sectors
as for some prescriptions and values of $\varepsilon$ the validity of the inequalities 
\eqref{eq:eigenf-neq} and \eqref{eq:eigenf-neq} (generalized to include the initial 
data at two points) as well as the statement of nonvanishing of $|c_1|+|c_2|$ might be affected 
near $v=0$ by the different behavior of the functions $f_{\pm},f_o,B$ there. 
Therefore we cannot exclude the existence of normalizable eigenfunction in those 
cases. One can see however, that, as up to the transformation \eqref{eq:super-trans} 
the eigenfunctions have the same asymptotic properties as the (corresponding to the 
negative eigenvalues) eigenfunctions of $\Theta_{\Lambda}$ for $\Lambda\leq 0$. 
In consequence all the normalizable eigenfunctions have to decay exponentially 
($\Lambda>\Lambda_c$) or like $O(v^{-3/2}\ln(v))$ ($\Lambda=\Lambda_c$). Furthermore, due to the form of
coefficients of $\Theta_{\Lambda}$ \eqref{eq:ev2-gen} they have to enter this behavior
already at $|v| \approx 4$. This is possible only for low values of $\omega$ as for 
the larger ones the term $\omega^2$ is a dominating one at $|v|\leq 4$, which again 
forces the behavior similar to the asymptotic one. In consequence any possible 
normalizable eigenfunctions necessarily correspond to small eigenvalues and are peaked near 
the classical singularity.

\section{\label{sec:concl}Conclusions}

We have considered the evolution operator $\Theta_{\Lambda}$ defining the evolution 
of the isotropic flat universe with massless scalar field and positive cosmological 
constant quantized within the framework of Loop Quantum Cosmology. For the investigation
three exact forms of the operator corresponding to particular prescriptions \cite{aps-imp}
(APS), \cite{acs} (sLQC) and \cite{mmo} (MMO) were selected. Our main goal was the 
verification of its self-adjointness as the condition necessary to generate a unique 
unitary evolution in the Schr\"odinger picture. We also investigated the properties of
the Hilbert spaces defined by the spectra of possible self-adjoint extensions of 
$\Theta_{\Lambda}$.

The results of the studies happen to depend on the value of the cosmological constant 
$\Lambda$. Namely, one can divide the set of its values onto two regions separated 
by the critical value related with the critical energy density \cite{aps-imp} via the equality 
$\Lambda_c = 8\pi G\rho_c$ for which the properties of $\Theta_{\Lambda}$ are 
qualitatively different.

For $0<\Lambda<\Lambda_c$ (denoted as subcritical) $\Theta_{\Lambda}$
admits many self-adjoint extensions, each of them defining inequivalent (at least 
at the mathematical level) unitary evolution. The extensions are labeled by the 
elements of the $U(1)$ or $U(2)$ group, depending on the superselection sector. 
In particular, once the studies are restricted to the symmetric functions only, the 
groups of labels $G$ are 
\begin{equation}
  G = \begin{cases}
         U(1) , & \MMO, \\
         U(2) , &\APS\text{ and }\sLQC,\, \varepsilon\neq 0,2 , \\
         U(1) , &\APS\text{ and }\sLQC,\, \varepsilon= 0,2 .
      \end{cases}
\end{equation}

Once the self-adjoint extensions were identified their spectral properties were 
also studied. It was shown that the spectrum of each extension is discrete, thus 
the physical Hilbert space spanned by the set of normalizable eigenfunctions of
$\Theta_{\Lambda}$.

For $\Lambda\geq\Lambda_c$ (supercritical) found relation with the operators 
$\Theta_{\Lambda}$ for $\Lambda\leq 0$ allowed to show, that the evolution operator
is essentially self-adjoint, thus generating a unique unitary evolution. Further studies
have shown however that for the superselection sector $\varepsilon=0$ the positive part 
of the spectrum of $\Theta_{\Lambda}$ is empty, thus the physical Hilbert space defined 
by it is trivial. This situation might change in other superselection sectors. There 
however, even if nontrivial, the Hilbert space does not admit any physically 
interesting states.

The result of the above paragraph is analogous to the properties of the scalar field energy density 
operator performed in \cite{klp-aspects}, where it was shown, that for $\Lambda\geq\Lambda_c$
the absolutely continuous part of the spectrum of that operator is entirely non-positive 
and the eigenfunctions corresponding to the positive elements of the spectrum (necessarily 
belonging to its discrete part) are peaked about $v=0$.

In the subcritical case $\Lambda<\Lambda_c$ the existence of non-unique extensions 
implies in particular that the quantum evolution of the system is not explicitly unique. However 
the detailed studies of its dynamics show \cite{ap-posL} that in the semiclassical 
regime the dynamical predictions are surprisingly unique and in the limit $v\to\infty$
consistent with the (unique) analytic extension of the classical trajectory.

To conclude, let us note that the results regarding self-adjointness directly extend
(with the exception of the case $\Lambda=\Lambda_c$) to the cases of different topologies 
($K=\pm 1$), as the terms in $\Theta_{\Lambda}$ present in such models are subleading 
with respect to the term $\Lambda V(v)$. Furthermore applying the methods presented 
here one can easily show that for the models $\Lambda=0$ and $K=-1$ (defined in \cite{hiperb}) 
the evolution operator also admits non-unique extensions. By the relation \eqref{eq:super-trans} 
this result applies also to $\Lambda=\Lambda_c, K=+1$. On the other hand the same argument and 
\cite{kls-closed} imply the self-adjointness for $\Lambda=\Lambda_c, K=-1$.

\begin{acknowledgments}
  We would like to thank Abhay Ashtekar and Jerzy Lewandowski for extensive discussions, 
  and to Guillermo Mena Marug\'an for carefully reading the draft and helpful comments.
  The work was supported in part by the Spanish MICINN project no FIS2008-06078-C0303,
  the Polish Ministerstwo Nauki i Szkolnictwa Wy\.zszego grants 1~P03B~075~29, 182/N-QGG/2008/0
  and the Foundation for Polish Science grant Master. TP acknowledges also the financial aid
  by the I3P program of CSIC and the European Social Fund and the funds of the European 
  Research Council under short visit grants 2127 and 3024 of QG network.
\end{acknowledgments}


\begin{thebibliography}{99}
\bibitem{argap}
    The quantity was originally introduced in \cite{aps-imp} and identified with
    the $1$-st nonzero eigenvalue of the LQG area operator \cite{lqg-th} 
    $\Delta=2\sqrt{3}\pi\gamma\lPl^2$, however more careful analysis of the relation 
    between the LQG graphs and the objects defined in LQC \cite{awe-entr} indicates 
    that this value should be in fact twice larger. The numerical value of $\Lambda_c$
    in the abstract was computed exactly for the modified value of $\Delta$ (and the value
    of $\gamma$ found in \cite{BI-value}) via \eqref{eq:Lambda-c}.
\bibitem{rep-diff} In comparizon to \cite{aps-imp} here we use a slightly different 
    representation obtained from the one of APS via the unitary mapping $\phi(v)
    \mapsto [B(v)]^{1/2}\phi(v)$.
\bibitem{GaSch-note} When comparing this claim with the results of \cite{klp-ga}
    one might find an apparent contradiction. It is not however the case, as the 
    exact forms of the constraint operator considered there and in this article 
    are different.
\bibitem{cont-note} In the case $\Lambda=\Lambda_c$ the spectrum may contain one element 
    of the absolutely continuous spectrum, namely $0$. However, since it is just a set 
    of measure zero, we can safely neglect it.
\end{thebibliography}
\end{document}